\documentclass[epj, nopacs]{svjour}
\usepackage{latexsym}
\usepackage{graphicx}
\usepackage{amssymb}
\usepackage{amsmath}
\usepackage{bm}
\usepackage{epsfig}
\usepackage{wasysym}
\usepackage{soul}
\usepackage{euscript,stmaryrd}
\usepackage{xcolor}

\begin{document}

\title{Explosive percolation on the Bethe lattice is ordinary}
\titlerunning{Explosive percolation on the Bethe lattice is ordinary}

\author{Young Sul Cho
\thanks{e-mail: yscho@jbnu.ac.kr}
}                   

\authorrunning{Young Sul Cho}

%
%
\institute{Department of Physics, Jeonbuk National University, Jeonju 54896, Korea \and Research Institute of Physics and Chemistry, Jeonbuk National University, Jeonju 54896, Korea}
\date{Received: date / Revised version: date}
%
\abstract{
The Achlioptas process, which suppresses the aggregation of large-sized clusters, can exhibit an explosive percolation (EP) where the order parameter emerges abruptly yet continuously in the thermodynamic limit. It is known that EP is accompanied by an abnormally small critical exponent of the order parameter. 
In this paper, we report that a novel type of EP occurs on a Bethe lattice, where the critical exponent of the order parameter is the same as in ordinary bond percolation based on numerical analysis. This is likely due to the property of a finite Bethe lattice that the number of sites on the surface with only one neighbor is extensive to the system size. To overcome this finite size effect, we consider an approximate size of the cluster that each site on the surface along its branch belongs to, and accordingly approximate the sizes of an extensive number of clusters during simulation. As a result, the Achlioptas process becomes ineffective and the order parameter behaves like that of ordinary percolation at the threshold. We support this result by measuring other critical exponents as well.
\PACS{
      {PACS-key}{discribing text of that key}   \and
      {PACS-key}{discribing text of that key}
     } 
} 

\maketitle

\section{Introduction}
\label{sec:intro}

Percolation is a phenomenon where fluid penetrates through a porous material and percolates to the other side~\cite{stauffer}, as seen in examples such as the sol--gel transition~\cite{flory1} and the metal--insulator transition~\cite{con_insul,last:1971}. To understand this phenomenon theoretically, ordinary bond percolation has been studied on various lattices~\cite{kim_percolation,saberi_bethe}.

In bond percolation on a lattice of dimension $d$, each bond is occupied with a probability of $p$. As a result, a continuous transition occurs for $d \geq 2$ with the critical exponent $\beta$ as  
\begin{equation}
P_{\infty} \propto (p-p_c)^{\beta}~~\text{as}~~p \rightarrow p^+_c,
\label{eq:beta}
\end{equation}
where the order parameter $P_{\infty}$ is the probability that a randomly selected site in the lattice is connected to the infinite cluster along occupied bonds. It is known that $\beta=5/36$ for $d=2$, and $\beta$ increases with $d$ up to $\beta=1$ for $d=6$. Then, $\beta = 1$ for $d \geq 6$. Among various lattice models, the Bethe lattice corresponds to an infinite dimensional lattice~\cite{er}, and bond percolation on the Bethe lattice exhibits a continuous transition with $\beta = 1$~\cite{Bethe_bond}.

The Achlioptas process refers to the occupation of a bond among several bond candidates following a given rule to suppress the aggregation of large clusters~\cite{Achlioptas:2009}. When bonds are occupied in a lattice under the Achlioptas process, $P_{\infty}$ increases abruptly yet continuously for $p \geq p_c$ as Eq.~(\ref{eq:beta}) in the thermodynamic limit. Due to such a behavior, the percolation model that the Achlioptas process applies to is called the explosive percolation (EP) model~\cite{Achlioptas:2009}. The EP model has been studied on lattices of dimensions $d \geq 2$~\cite{ziff_prl,cho_science}, and it has been reported that $\beta \ll 1$ irrespective of dimensions~\cite{ziff_lattice:2010,filippo_pre:2010,fss_exp,local,tricritical,hklee,choi,dacosta_prl,dacosta_pre,grassberger,riordan,smoh:2016,eppre:2023,epprl:2023,souza_nphy,cho_kahng_review,Jan_gap1,Jan_gap2,saberi_review}.

In~\cite{EP_Bethe}, the EP model was studied on the Bethe lattice,
showing that $P_{\infty}$ increases abruptly at the threshold as in a usual EP model.
However, in this previous study, the critical exponents were not investigated.
In the present paper, we first modify the model to remove invalid assumptions and then obtain critical exponent values.
Surprisingly, we find that the critical exponents $\beta$ have the same values as those of bond percolation on the Bethe lattice, which means that the model indeed exhibits a novel type of EP. We support this claim by showing that some other critical exponents $\gamma, \gamma', \tau$, and $\sigma$ also have the same values as those of bond percolation. We confirm that this phenomenon is not caused by the modification of the model by observing the same phenomenon in the original model.

\begin{figure}[t!]
\includegraphics[width=1.0\linewidth]{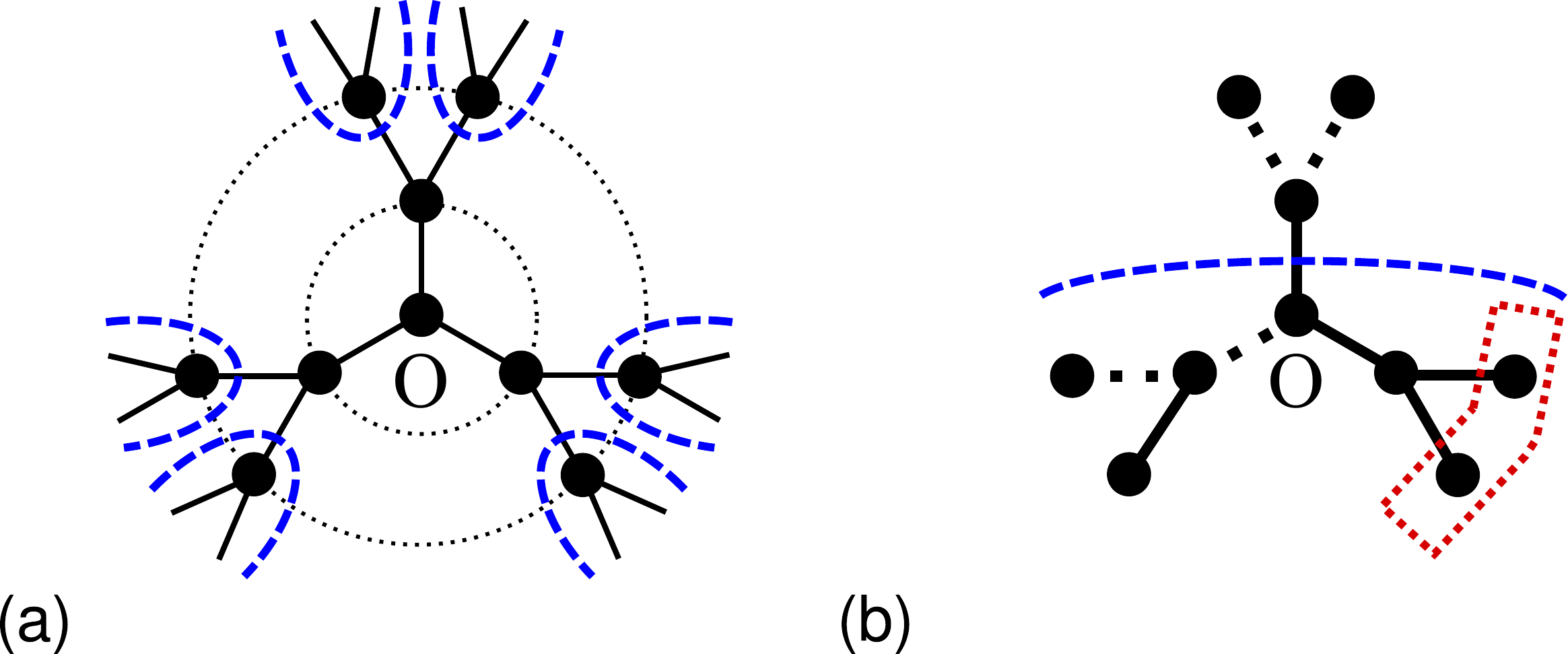}
\caption{(a) Schematic diagram of the Bethe lattice with $z=3$ composed of a Cayley tree with radius $n=2$ and $z(z-1)^{n-1}=6$ infinite branches (dashed lines). 
(b) In a Cayley tree composed of five occupied bonds, a cluster connected along randomly selected $z-1=2$ directions from \textbf{O} (below the dashed line) contains $s=4$ sites including $t=2$ sites on the surface (surrounded by the dotted line).}
\label{Fig:Bethe_schematic}
\end{figure}

\begin{figure}[t!]
\includegraphics[width=1.0\linewidth]{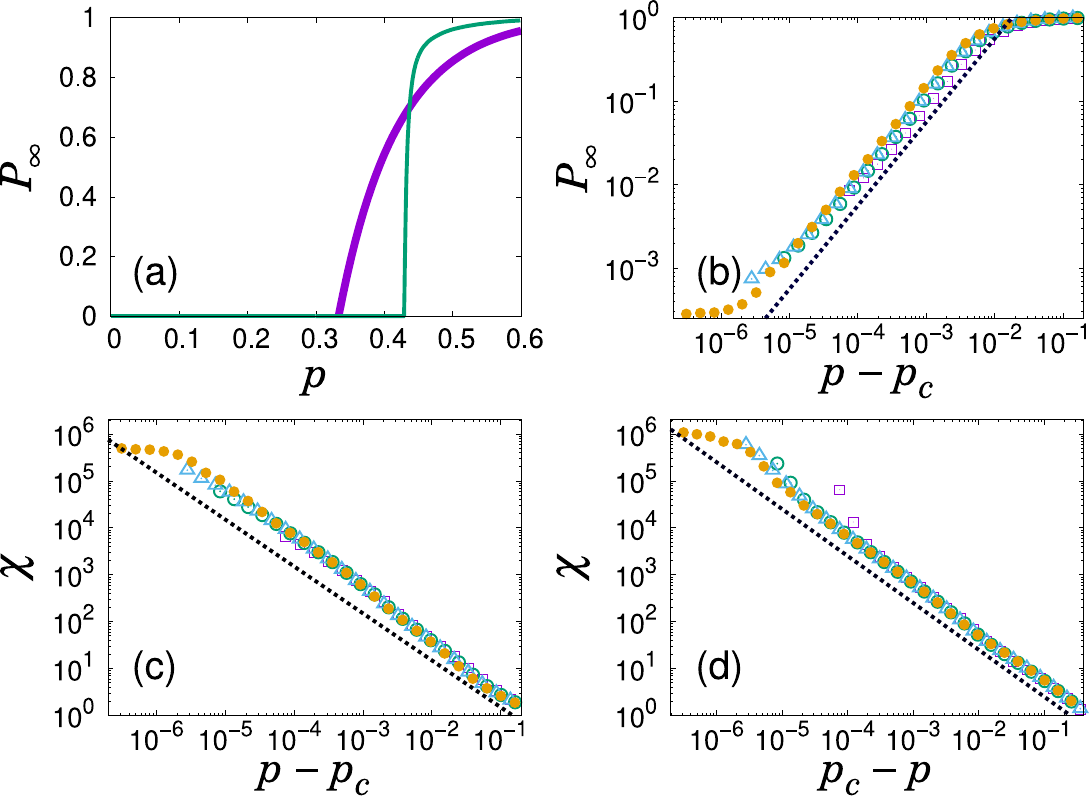}
\caption{(a) Theoretical result of bond percolation (thick purple line) and simulation result of EP using $n=13$ (thin green line) on the Bethe lattice with $z=4$. 
(b--d) Simulation results of EP
using $n=8$ ($\square$), $10$ ($\circ$), $11$ ($\triangle$), and $13$ ($\bullet$) with $z=4$, where the slopes of the dotted lines are (b) $1$, (c) $-1$, and (d) $-1$.} 
\label{Fig:Qnst_Pinf_Chi}
\end{figure}

This paper is organized as follows. In Sec.~\ref{sec:epmodel}, we introduce a modified model that corrects some invalid assumptions used in the original work~\cite{EP_Bethe}.
In Sec.~\ref{sec:criticalexponent}, we show that the measured critical exponent values in the modified model are the same as those in bond percolation.
In Sec.~\ref{sec:discussion}, we conclude by discussing the results.
In the Appendix, we provide details on how some assumptions used in the original model are not valid. We also show that the measured critical exponent values in the original model are the same as those in bond percolation, which confirms that our results hold regardless of the model.

\section{Modification of the explosive percolation model on the Bethe lattice}
\label{sec:epmodel}

In~\cite{EP_Bethe}, self-consistent equations for arbitrary percolation on the Bethe lattice were established under the assumption that recursive behavior is observed in the branches of an internal Cayley tree with an arbitrary radius $n' \leq n$. By solving these equations for an arbitrary pair $n' < n$ simultaneously, $P_{\infty}$ can be obtained. However, this assumption is actually not valid for the EP model, and thus we modify the self-consistent equations not to introduce $n'$.

A Bethe lattice is a tree where each site has $z$ neighbors~\cite{kim_percolation,Bethe_bond}. 
We consider a Cayley tree with radius $n$, 
where the total number of sites within the Cayley tree is given by $N = 1+z((z-1)^n-1)/(z-2)$. 
A Bethe lattice can then be treated as a Cayley tree where each site on the surface (outermost shell) is the root of an infinite branch [Fig.~\ref{Fig:Bethe_schematic}(a)].
If we treat a randomly selected site in the Bethe lattice as the center of a Cayley tree called \textbf{O}, then
$P_{\infty}$ is the probability that \textbf{O} is connected to the infinite cluster along occupied bonds.

To obtain $P_{\infty}$ on the Bethe lattice when $p$ fraction of bonds are occupied following a given rule, we occupy bonds on the Cayley tree with radius $n$ following the rule and regard the fraction of occupied bonds on the tree as $p$. Then, $P_{\infty}$ can be calculated using $A$ and $P_{nst}$ as
\begin{equation}
P_{\infty} = 1-\sum_{s,t}P_{nst}(1-A)^t,
\label{eq:P_inf}
\end{equation}
where $A$ is the probability that each site on the surface belongs to the infinite cluster along its branch and
$P_{nst}$ is the probability that the cluster to which \textbf{O} belongs has $s$ sites with $t$ of those being on the surface~\cite{EP_Bethe}.

Due to the recursive structure of the Bethe lattice, we assume that the probability that \textbf{O} is connected to the infinite cluster along randomly selected $z-1$ directions would also be $A$ for a sufficiently large $n$. Then $A$ 
is obtained by
\begin{equation}
A = 1-\sum_{s,t} Q_{nst}(1-A)^t,
\label{eq:A_recursive}
\end{equation}
where $Q_{nst}$ is the probability that a
cluster connected along randomly selected $z-1$ directions from \textbf{O} contains $s$ sites including $t$ sites on the surface [Fig.~\ref{Fig:Bethe_schematic}(b)]. 

We note that the above assumption is valid for any $n \geq 1$ in the bond percolation. Considering Eqs.~(\ref{eq:P_inf}) and (\ref{eq:A_recursive}) for $n = 1$ with $P_{1t}={z \choose t}p^t(1-p)^{z-t}$ and $Q_{1t}={z-1 \choose t}p^t(1-p)^{z-1-t}$, we reproduce the results $P_{\infty}=1-(1-pA)^z$ and $A=1-(1-pA)^{z-1}$ in the bond percolation.

We now apply an Achlioptas process called the product rule~\cite{Achlioptas:2009} to the Bethe lattice. When the product rule is applied to a finite lattice, at each time step, two unoccupied bond candidates are chosen randomly, and among the two the bond with a smaller product of connecting cluster sizes is occupied. In this manner, the emergence of an infinite cluster is suppressed and $P_{\infty}$ increases rapidly at the delayed threshold. In particular, the product rule in a finite Bethe lattice requires $S_b$, which is the average size of the cluster in which each site on the surface is connected along its branch (toward the outside of the lattice) in order to suppress each site from being connected to an infinite cluster through its branch.

Similar to the assumption used to derive Eq.~(\ref{eq:A_recursive}), we assume that the average size of the cluster containing \textbf{O} connected to the infinite cluster along randomly selected $z-1$ directions would also be $S_b$ for a sufficiently large $n$. Then $S_b$ is obtained by
\begin{equation}
S_b = \sum_{s,t}(s-t+tS_b)Q_{nst}(1-A)^{t-1},
\label{eq:Sb_recursive}
\end{equation}
where $(1-A)^{-1}$ on the right side is the normalization factor of $Q_{nst}(1-A)^t$ given by Eq.~(\ref{eq:A_recursive}).

In principle, $P_{\infty}, A, S_b$ of the EP model on the Bethe lattice are obtained by solving Eqs.~(\ref{eq:P_inf})--(\ref{eq:Sb_recursive}) computationally as follows. At the beginning of the iteration, $A^{(1)}(p)=0$ and $S^{(1)}_b(p)=1$ are given. 
In the $i$-th iteration, each simulation of occupying bonds from $p=0$ to $p=1$ following a given rule using $A^{(i)}(p)$ and $S^{(i)}(p)$ is repeated many times.
Then $Q^{(i)}_{nst}(p)$ and $P^{(i)}_{nst}(p)$ are obtained by investigating at what fraction of simulations the cluster including \textbf{O} has $s,t$ values for each.
Finally, $P_{\infty}^{(i+1)}, A^{(i+1)}, S_b^{(i+1)}$ for the $(i+1)$-th iteration are obtained by solving Eqs.~(\ref{eq:P_inf})--(\ref{eq:Sb_recursive}) after substituting $P_{nst}=P^{(i)}_{nst}$ and $Q_{nst}=Q^{(i)}_{nst}$ into those equations.
This iteration continues until $A^{(i)}, S_b^{(i)}, P_{\infty}^{(i)}$ saturate to $A, S_b, P_{\infty}$, satisfying Eqs.~(\ref{eq:P_inf})--(\ref{eq:Sb_recursive})
approximately.

For each simulation of the $i$-th iteration, we begin with $p=0$ where all bonds on the Cayley tree are unoccupied. 
When a bond is occupied, $p$ increases by $1/(N-1)$ because the total number of bonds on the Cayley tree is $N-1$. 
Here, the product rule is applied and the bond to be occupied at $p$ is selected according to the following steps.
\begin{itemize}
\item[(i)] Two unoccupied bonds $b_1$ and $b_2$ are chosen randomly, where the two sites at ends of each bond $b_{\ell}$ are named 
$b_{\ell 1}$ and $b_{\ell 2}$.
\item[(ii)] The total number of sites included in the cluster that $b_{\ell k}$ belongs to is given by $s_{\ell k}$, 
and among those, the number of sites on the surface is given by $t_{\ell k}$.
\item[(iii)] The cluster $b_{\ell k}$ belongs to is determined to be a finite cluster independently at random with probability $(1-A^{(i)}(p))^{t_{\ell k}}$.
\item[(iv)] If the cluster that $b_{\ell k}$ belongs to is a finite cluster, its size $Z_{\ell k}$ is given by $Z_{\ell k}=s_{\ell k}+t_{\ell k}\big[S^{(i)}_b(p)-1\big]$, and otherwise $Z_{\ell k}$ is infinite.
\item[(v)] The bond with the smaller product $Z_{\ell 1}Z_{\ell 2}$ among the two bonds is occupied. At this time, if both products are infinite, one of the two bonds is occupied randomly.
\end{itemize}
Each simulation ends when $N-1$ links are occupied in this manner.

To obtain all data in the figures, we perform at least $10^5$ simulation runs in each iteration and average over $10^2$ iterations after saturation.

In Fig.~\ref{Fig:Qnst_Pinf_Chi}(a), the result of $P_{\infty}(p)$ of the product rule on the Bethe lattice is depicted. As in a usual EP model, $P_{\infty}$ increases rapidly at the delayed threshold compared to that of bond percolation.

\section{Observation of the same critical exponent values as in ordinary bond percolation}
\label{sec:criticalexponent}

\begin{figure}[t!]
\includegraphics[width=1.0\linewidth]{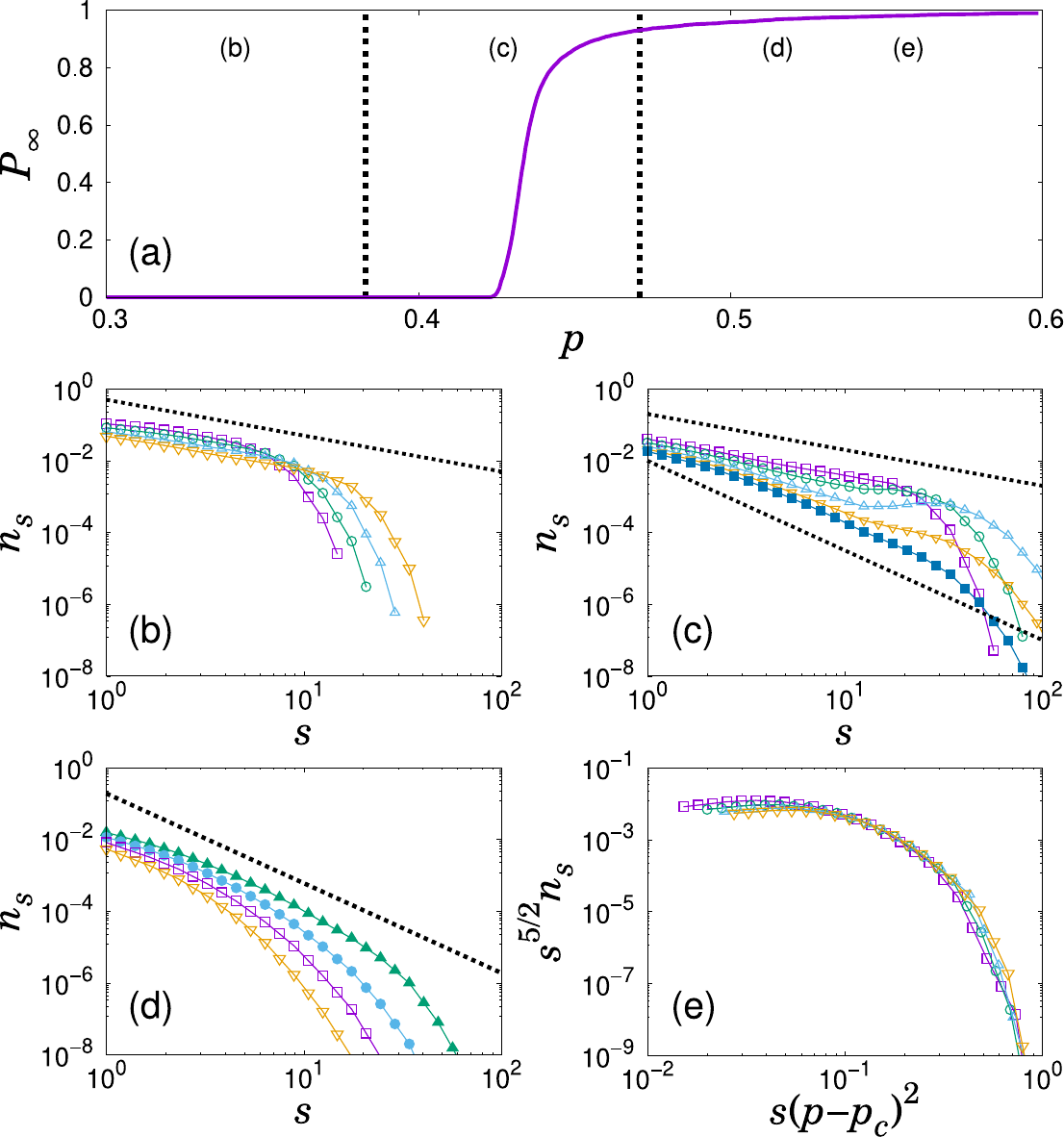}
\caption{Simulation result of the EP model on the Bethe lattice using $n=13$ and $z=4$. (a) Each range of $p$, separated by dotted lines, contains the $p$ values used in the panel with the corresponding label shown above. (b) $n_s$ for $p=0.30$ ($\square$), $0.33$ ($\circ$), $0.35$ ($\triangle$), and $0.37$ ($\triangledown$). The slope of the dotted line is $-1$. 
(c) $n_s$ for $p=0.39$ ($\square$), $0.41$ ($\circ$), $0.43$ ($\triangle$), $0.44$ ($\triangledown$), and $0.46$ ($\blacksquare$). The slopes of the dotted lines are $-1$ (upper) and $-5/2$ (lower). 
(d) $n_s$ for $p=0.48$ ($\blacktriangle$), $0.52$ ($\bullet$), $0.55$ ($\square$), and $0.60$ ($\triangledown$). The slope of the dotted line is $-5/2$.
(e) Data collapse of $n_s$ for $p=0.55$ $(\square)$, $0.57$ $(\circ)$, $0.59$ $(\triangle)$, and $0.60$ $(\triangledown)$
using $p_c=0.4313$, which is the estimated value of $p_c$
in the thermodynamic limit $N \rightarrow \infty$.} 
\label{Fig:Pinf_cdist}
\end{figure}

In this section, we numerically estimate four percolation critical exponents in the EP model on the Bethe lattice as $\{\beta=1, \gamma=\gamma'=1, \tau=5/2, \sigma=1/2\}$, 
which are those of bond percolation.
Based on the previous study~\cite{EP_Bethe}, $P_{\infty}$ would exhibit a continuous transition of the form in Eq.~(\ref{eq:beta}).

Using the Cayley tree of different sizes $n$ for a fixed $z$, we obtain $\beta \approx 1$ as shown in Fig.~\ref{Fig:Qnst_Pinf_Chi}(b).
To draw this figure, we define $p^{(i)}_c = \text{min}\{p|P^{(i)}_{\infty}(p)>0\}$ and plot $\langle P^{(i)}_{\infty}(p_c^{(i)}+\overline{p}) \rangle$ with respect to $\overline{p}>0$, where $\langle \cdot \rangle$ denotes the average over different iterations indexed by $i$ after saturation.
In this manner, the average clearly reflects the shape of $P^{(i)}_{\infty}(p) \propto (p-p^{(i)}_c)$ as $p \rightarrow p^{(i)+}_c$ in each individual iteration.

The average size of the finite cluster that \textbf{O} belongs to is given by
\begin{equation}
\chi = \frac{\sum_{s,t}(s-t+tS_b)P_{nst}(1-A)^t}{1-P_{\infty}},
\label{eq:chi}
\end{equation}
where $(1-P_{\infty})^{-1}$ is the normalization factor of $P_{nst}(1-A)^t$ given by Eq.~(\ref{eq:P_inf})~\cite{EP_Bethe}. This corresponds to the average size of the finite cluster to which a randomly selected site belongs in a percolation model. Therefore, the critical behavior of $\chi$ is given by
\begin{equation}
\chi \propto
\begin{cases}
    (p-p_c)^{-\gamma} ~~\text{as}~~ p \rightarrow p^+_c, \\
    (p_c-p)^{-\gamma'} ~~\text{as}~~ p \rightarrow p^-_c.
\end{cases}
\end{equation}
Using the Cayley tree of different sizes $n$ for a fixed $z$, we obtain $\gamma \approx 1$ and $\gamma' \approx 1$ as shown in Fig.~\ref{Fig:Qnst_Pinf_Chi}(c) and (d), respectively. 
To draw these figures, $\langle \chi(p_c^{(i)}+\overline{p}) \rangle$ is plotted with respect to $\overline{p}$ and $-\overline{p}$ in Fig.~\ref{Fig:Qnst_Pinf_Chi}(c) and (d), respectively.

The other two exponents, $\{\sigma, \tau\}$, describe the scaling behaviors of the cluster size distribution $n_s$ as $n_s=s^{-\tau}f(s|p-p_c|^{1/\sigma})$,
where $n_s$ is the probability of forming a cluster of size $s$. Therefore, $sn_s$ is the probability that a randomly selected site belongs to a cluster of size $s$.

In the EP model on the Bethe lattice, $sn_s$ is obtained using the relation $sn_s = \sum_{t} P_{nst}(1-A)^{t}$ after treating \textbf{O} as a randomly selected site. However, $sn_s$ obtained using this relation is not exact when $p$ is sufficiently close to $p_c$, such that the correlation length is longer than the finite lattice radius $n$. Although such finite size effect is a common property of a lattice, $N$ increases exponentially with the radius in the Bethe lattice, unlike how $N$ usually increases polynomially with the radius in other lattice types. As a result, the radius of the Bethe lattice is short in the range of $N$ performed in simulation, and thus the range of $p$ near $p_c$ where the relation $sn_s = \sum_{t} P_{nst}(1-A)^{t}$ does not hold is wide, about $0.1$, as can be seen from the description of Fig.~\ref{Fig:Pinf_cdist} in the following paragraph. This abnormal property of the Bethe lattice was also confirmed by comparing with theoretical results in bond percolation (not shown here).

In Fig.~\ref{Fig:Pinf_cdist}, $n_s(p)$ was obtained using the relation $n_s(p) = (1/s)\sum_t P_{nst}(p)$ for various $p$ values in the largest Bethe lattice we computed, where $P_{nst}(p) = \langle P^{(i)}_{nst}(p)\rangle$. 
At first, the value of $p_c$ in the thermodynamic limit $N \rightarrow \infty$ 
is extrapolated by investigating the $N$ dependence of $\langle p^{(i)}_c \rangle$. When $p$ is sufficiently smaller than $p_c$, then $n_s \propto s^{-1}$ and the range of the power-law regime increases with $p$ as shown in Fig.~\ref{Fig:Pinf_cdist}(b).  
However, $\tau$ should be larger than 2. Otherwise, $n_s$ for all $s$ shrink to zero as $p \rightarrow p_c$, which is contradictory to $p_c < 1$.
As a result, $\tau$ increases rapidly as $p$ approaches $p_c$, and $\tau = 5/2$ for $p > p_c$
as shown in Fig.~\ref{Fig:Pinf_cdist}(c). 
Moreover, a bump in $n_s$ exists as clusters of similar sizes frequently form in the large $s$ range. This phenomenon occurs because the clusters cannot extend outside of the Cayley tree even though the correlation length is longer than the radius. Therefore, $n_s$ is not exact due to the finite size effect of the Cayley tree in the range of $p$.

As $p$ is sufficiently larger than $p_c$, the bump disappears and $n_s$ follows the conventional scaling behavior of bond percolation~\cite{friedman,hooyberghs,cho_scirep,choprl:2016,park_hybrid,choi_hybrid},
as shown in Fig.~\ref{Fig:Pinf_cdist}(d) and (e).  
In conclusion, we claim that $n_s$ is given by 
\begin{equation}
n_s = s^{-\tau}f\big[s(p-p_c)^{1/\sigma}\big]~~\text{for}~~p > p_c,
\end{equation}
with $\tau=5/2$ and $\sigma = 1/2$ for the scaling function $f(x)$, which is flat for $x \ll 1$ but decays rapidly for $x \gg 1$~\cite{stauffer}.

For $p < p_c$, $\tau$ seems to be smaller than $5/2$ by the suppression of infinite cluster emergence in the EP model. However, it increases rapidly to $5/2$ as $p \rightarrow p_c^-$.
Therefore, further investigation is needed to determine whether $\tau=5/2$ and $\sigma=1/2$ in the subcritical region.

\section{Discussion}
\label{sec:discussion}

In this paper, we investigated four percolation critical exponents in the EP model on the Bethe lattice. As a result, we obtained $\{\beta \approx 1, \gamma \approx 1, \tau \approx 5/2, \sigma \approx 1/2\}$ in the supercritical region $p>p_c$ and $\gamma' \approx 1$ in the subcritical region $p<p_c$.
These critical exponent values are consistent with those in bond percolation, and thus the possibility exists that the two models belong to the same universality class. To confirm this, it is necessary to examine additional critical exponent values in EP on the Bethe lattice.

Finally, we briefly discuss why $\beta \approx 1$ in this model in contrast to $\beta \ll 1$ in other EP models. For comparison, we consider the EP model on finite-dimensional hypercubic lattices of dimension $d$. 
We obtain the average value of $P_{\infty}(p)$ by occupying $p$ fraction of bonds following the product rule repeatedly in a finite lattice and measure $\beta$ by increasing the lattice size. In this case, all sites have $2d$ number of neighbors in the thermodynamic limit. Therefore, one does not need to consider an approximate size of the clusters outside of the system, such as $S_b$, and exact sizes of all clusters are given at each $p$ during simulation. We thus expect that the product rule, which requires a size comparison between clusters, is effectively implemented, resulting in $\beta \ll 1$.

On the other hand, in the case of the Bethe lattice, it is not possible to obtain $\beta$ by increasing the radius of the Cayley tree where $P_{\infty}$ is calculated under the product rule. 
This is because the number of neighbors of each site is consistently $z$ on the Bethe lattice, but each site on the surface has only one neighbor and the fraction of these sites remains finite even in the thermodynamic limit on the Cayley tree, contrary to the finite-dimensional hypercubic lattices. Thus, it is not possible to treat the Cayley tree as a Bethe lattice of finite size. We therefore have to consider an infinite branch attached to each site on the surface with an approximate size $S_b$ of the cluster along each branch in order to computationally handle the (infinite) Bethe lattice. Consequently, approximate sizes of an extensive number of clusters on the surface are given and the product rule would be ineffective, resulting in $\beta \approx 1$ like in bond percolation.

\section*{Acknowledgement}
This paper was supported by research funds of Jeonbuk National University in 2023. 
We thank Huiseung Chae for his valuable comments on the simulation method.

\section*{Authors contributions}
All authors contributed equally to the paper.

\section*{Appendix A: Observation of the same critical exponent values in the original model}
\label{sec:appendix}

In this Appendix, we show that $P_{\infty} \propto (p-p_c)^{\beta}$ as $p \rightarrow p^+_c$ and $\chi \propto |p-p_c|^{-\gamma (\gamma')}$ as $p \rightarrow p_c^+ (p_c^-)$, with $\beta \approx 1$ and $\gamma \approx \gamma' \approx 1$
reproduced in the original EP model on the Bethe lattice introduced in~\cite{EP_Bethe}.
Here, we do not investigate the $n_s$ of this model for simplicity.

We begin with the self-consistent equations introduced in the previous study. The previous study assumed that $P_{\infty}$ and $\chi$
obtained using $P_{n'st}$ of an internal Cayley tree with radius $n'$ for any $n' \leq n$ should be the same as those obtained using $P_{nst}$. Based on this assumption, two self-consistent equations were derived as follows.

To introduce the first equation, we denote $P_{\infty}$ obtained using an internal Cayley tree with the radius $n'$ as $P_{n' \infty}$, where it is given by $P_{n' \infty} = 1-\sum_{s,t}P_{n' st}(1-A)^t$. Then the above assumption implies that $P_{n\infty}=P_{n' \infty}=P_{\infty}$ for any $n' \leq n$, which leads to the first equation
\begin{equation}
\sum_{s,t}P_{nst}(1-A)^t = \sum_{s,t}P_{n'st}(1-A)^t
\tag{A1}
\label{eq:Pnst_Pn'st}
\end{equation}
for an arbitrary $n'< n$.

To introduce the second equation, we denote $\chi$ obtained using an internal Cayley tree with radius $n'$ as $\chi_{n'}$, where $\chi_{n'} = (1-P_{\infty})^{-1}\sum_{s,t}(s-t+tS_b)P_{n' st}(1-A)^t$.
Then the above assumption implies that $\chi_{n}=\chi_{n'}=\chi$ for any $n' \leq n$, which
leads to the second equation
\begin{align}
&\sum_{s,t}(s-t+tS_b)P_{nst}(1-A)^t \nonumber \\&=\sum_{s,t}(s-t+tS_b)P_{n'st}(1-A)^t
\tag{A2}
\label{eq:chin_chin'}
\end{align}
for an arbitrary $n'<n$.

In principle, Eqs.~(\ref{eq:Pnst_Pn'st}) and (\ref{eq:chin_chin'}) for a common $n' < n$ can be solved
by iterating until $A^{(i)}$ and $S_b^{(i)}$ converge to $A$ and $S_b$, thereby satisfying these equations following the process described in~\cite{EP_Bethe}.
Then $P_{\infty}$ and $\chi$ are obtained by substituting $A$ and $S_b$ into Eq.~(\ref{eq:P_inf}) and (\ref{eq:chi}).

\begin{figure}[t!]
\includegraphics[width=1.0\linewidth]{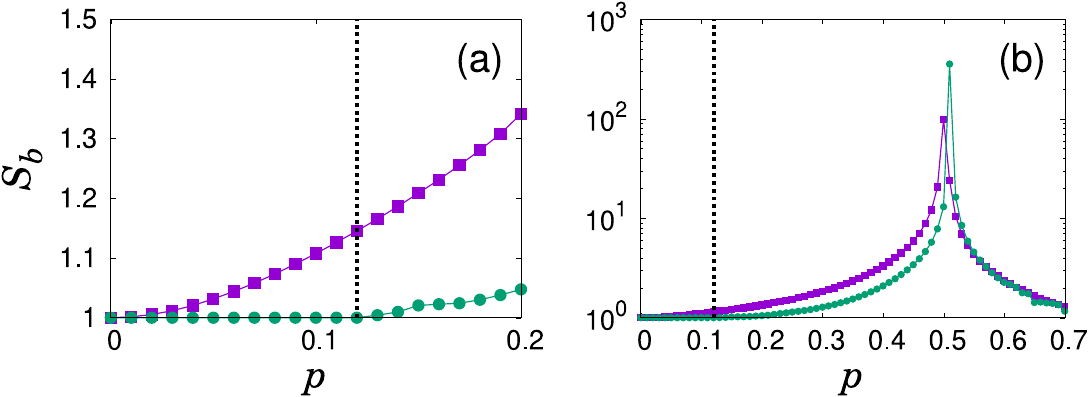}
\caption{Simulation result of the original EP model with $(n, n')=(10, 4)$ ($\bullet$) and $(n, n')=(10, 1)$ ($\blacksquare$). (a) $S_b$ for $(n,n')=(10,4)$ becomes nonzero as $p$ exceeds the dotted line. (b) $S_b$ in both conditions differs over the entire range of $p$.} 
\label{Fig:Pnst_problem}
\end{figure}

\begin{figure}[t!]
\includegraphics[width=1.0\linewidth]{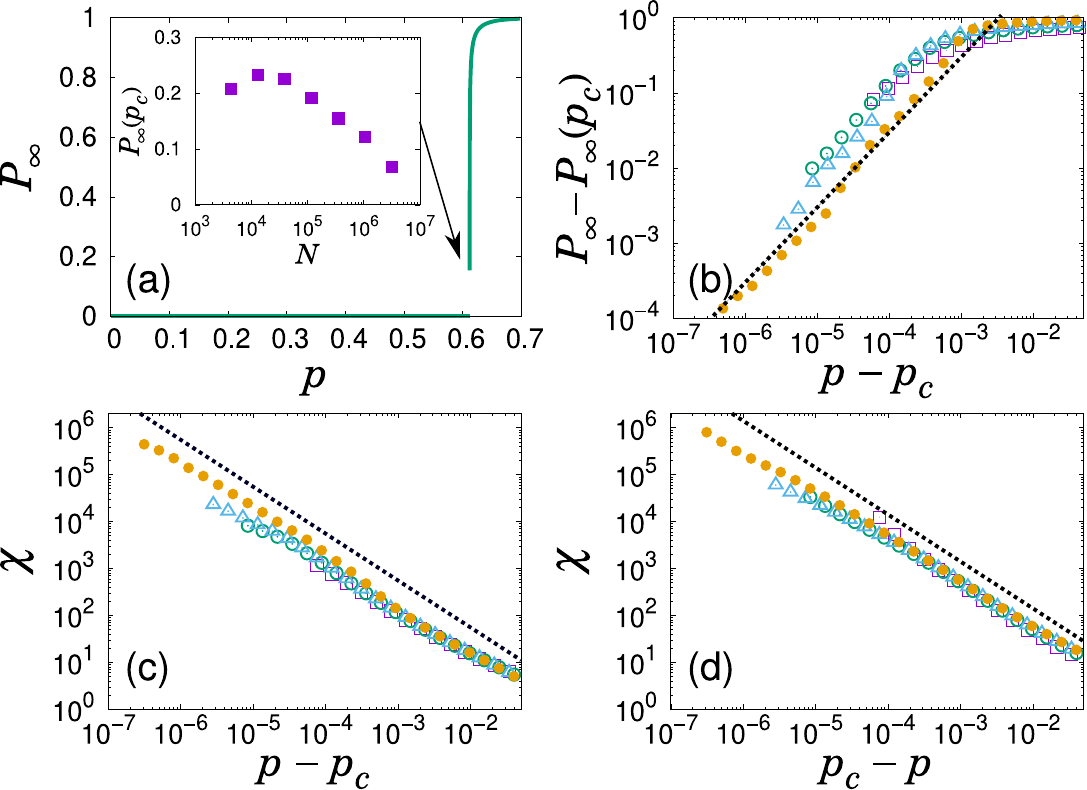}
\caption{Simulation result of the original EP model on the Bethe lattice with $z=4$. (a) $P_{\infty}(p)$ for $(n,n')=(11, 4)$. Inset: Jump size of $P_{\infty}$ at $p=p_c$ vs. $N$ for fixed $n'=4$. (b--d) Data obtained using $n=8$ ($\square$), $10$ ($\circ$), $11$ ($\triangle$), and $13$ ($\bullet$) for fixed $n'=4$. The slopes of the dotted lines are (b) $1$, (c) $-1$, and (d) $-1$.}
\label{Fig:nprime_Pinf_Chi}
\end{figure}

However, we find that $S_b$ depends on $n'$ for a fixed $n$, as shown in Fig.~\ref{Fig:Pnst_problem}, which violates the assumption that $S_b$ is the same irrespective of $n'$ for a fixed $n$. 
We can understand this as follows. At first, we consider a small $p$ range where $A=0$ and $P_{nst}=0$ for all $t \geq 1$. In this case, Eq.~(\ref{eq:chin_chin'}) is reduced to $\sum_{s}sP_{ns0} = \sum_{s,t}(s-t+tS_b)P_{n'st}$. Accordingly, there must be at least one $t>0$ satisfying $P_{n'st}>0$ in order for $S_b > 1$. 
As $n'$ becomes larger, the average size of the cluster to which \textbf{O} belongs must be larger to satisfy this condition.
Therefore, the initial range of $p$ where $S_b=1$ increases with $n'$, as shown in Fig.~\ref{Fig:Pnst_problem}(a). This initial difference between $S_b$ for different $n'$ leads to a difference between those in the entire range of $p$, as shown in Fig.~\ref{Fig:Pnst_problem}(b). Note that in the main text we suggested our method to avoid this dependence of $S_b$ on $n'$.

In Fig.~\ref{Fig:nprime_Pinf_Chi}, we show that the critical exponent values $\beta \approx 1$ and $\gamma \approx \gamma' \approx 1$ are also observed in the original EP model for the values of $(n, n')$ used in~\cite{EP_Bethe}.
As mentioned in~\cite{EP_Bethe}, the
average gap size $\langle P^{(i)}_{\infty}(p^{(i)}_c) \rangle > 0$ for a finite $n$ and $\langle P^{(i)}_{\infty}(p^{(i)}_c)\rangle \rightarrow 0$ as $n \rightarrow \infty$ 
[Fig.~\ref{Fig:nprime_Pinf_Chi}(a)]. To estimate $\beta$ in the thermodynamic limit $n \rightarrow \infty$, we measure
the slope of $\langle P^{(i)}_{\infty}(p) \rangle - \langle P^{(i)}_{\infty}(p^{(i)}_c)\rangle $ with respect to $(p-p_c)$,
where $p_c$ satisfies $\langle P^{(i)}_{\infty}(p_c) \rangle = \langle P^{(i)}_{\infty}(p^{(i)}_c)\rangle$. As shown in Fig.~\ref{Fig:nprime_Pinf_Chi}(b), the range of $p-p_c$ where the slope is close to 1 increases with $n$, which supports $\beta \approx 1$. Finally, $\gamma \approx 1$ and $\gamma' \approx 1$ are shown by plotting $\langle \chi(p_c^{(i)}+\overline{p}) \rangle$ with respect to $\overline{p}$ and $-\overline{p}$ in Fig.~\ref{Fig:nprime_Pinf_Chi}(c) and (d), respectively.

\end{document}